\documentstyle[preprint,aps,epsfig,tighten]{revtex}
\begin{document}
\draft
\preprint{
\parbox[t]{45mm} 
{MPG-VT-UR 214/01} 
}
\title{Chiral Lagrangian approach to the $J/\psi$ breakup cross section}

\author{V.~V.~Ivanov, Yu.~L.~Kalinovsky}
\address{Laboratory of Information Technologies, JINR Dubna, 141980 Dubna, 
Russia}
\author{D.~B.~Blaschke}
\address{Fachbereich Physik, Universit\"at Rostock, D-18051 Rostock, Germany\\
ECT*, Strada delle Tabarelle 286, I-38050 Villazzano (Trento), Italy\\
Bogoliubov Laboratory for Theoretical Physics, JINR Dubna, 
RU-141980 Dubna, Russia} 
\author{G.~R.~G.~Burau}
\address{Fachbereich Physik, Universit\"at Rostock, D-18051 Rostock, Germany} 
\maketitle
\begin{abstract}
We summarize the results of the $SU(4)$ chiral meson Lagrangian approach to 
the cross section for $J/\psi$ breakup by pion and rho meson impact and 
suggest a new scheme for the introduction of formfactors for the meson-meson
interaction.
This scheme respects the fact that on the quark level of description the
contact and the meson exchange diagrams are constructed by so-called 
box and triangle diagrams which contain a different number of vertex functions
for the quark-meson coupling.   
A model calculation for Gaussian vertex functions is presented and the 
relative importance of contact and meson exchange processes is discussed. 
We evaluate the dependence of the breakup cross section on the masses of the 
final D-meson states which
can be used for the study of in-medium effects on this quantity.
\vspace{5mm}
\noindent
\end{abstract}

\section{Introduction}
The $J/\psi$ meson plays a key role in the experimental search for the 
quark-gluon plasma (QGP) in heavy-ion collision experiments where an 
anomalous suppression of its production cross section relative to the 
Drell-Yan continuum as a function of the centrality of the collision has 
been found by the CERN-NA50 collaboration \cite{na50}. 
An effect like this has been predicted to signal QGP formation \cite{ms86} 
as a consequence of the screening of color charges in a plasma in close 
analogy to the Mott effect (metal-insulator transition) in dense electronic 
systems \cite{rr}. 
However, a necessary condition to explain $J/\psi$ suppression in the static 
screening model is that a sufficiently large fraction of $c\bar c$ pairs after 
their creation have to traverse regions of QGP where the temperature 
(resp. parton density) has to exceed the Mott temperature 
$T^{\rm Mott}_{{\rm J}/\psi}\sim 1.2 - 1.3~ T_c$ \cite{kms,rbs} for a 
sufficiently long time interval $\tau>\tau_{\rm f}$,  where 
$T_c\sim 170$ MeV is the critical phase transition temperature and 
$\tau_{\rm f}\sim 0.3 $ fm/c is the $J/\psi$ formation time. 
Within an alternative scenario \cite{rbs2}, $J/\psi$ suppression does not 
require temperatures well above the deconfinement one but can occur already 
at $T_c$ due to impact collisions by quarks from the thermal medium. 
An important ingredient for this scenario is the lowering of the reaction 
threshold for string-flip processes which lead to open-charm meson formation 
and thus to $J/\psi$ suppression. 
This process has an analogue in the hadronic world, where e.g. 
J/$\psi + \pi \rightarrow D^* + \bar D + h.c.$ could occur provided the 
reaction threshold of $\Delta E \sim 640$ MeV can be overcome by pion impact. 
It has been shown recently \cite{bbk} that this process and its in-medium 
modification can play a key role in the understanding of anomalous $J/\psi$ 
suppression as a deconfinement signal. 
Since at the deconfinement transition the $D$- mesons enter the continuum of 
unbound (but strongly correlated) quark- antiquark states (Mott- effect), the 
relevant threshold for charmonium breakup is lowered and the reaction rate for 
the process gets critically enhanced. Thus a process which is negligible in 
the vacuum may give rise to additional (anomalous) $J/\psi$ suppression when 
conditions of the chiral/ deconfinement transition and $D$- meson Mott effect 
are reached in a heavy-ion collision but the dissociation of the $J/\psi$ 
itself still needs impact to overcome the threshold which is still present but 
dramatically reduced. 
 
For this alternative scenario as outlined in \cite{bbk} to work the $J/\psi$ 
breakup cross section by pion impact is required and its dependence on the 
masses of the final state $D$- mesons has to be calculated. 
Both, nonrelativistic potential models \cite{mbq95,wsb00} and chiral 
Lagrangian models \cite{mm98,lk00,hg00} have been employed to determine the 
cross section in the vacuum. The results of the latter models appear to be 
strongly dependent on the choice of formfactors for the meson-meson vertices. 
This is considered as a basic flaw of these approaches which could only be 
overcome when a more fundamental approach, e.g. from a quark model, can 
determine these input quantities of the chiral Lagrangian approach. 
 
In the present paper we would like to reduce the uncertainties of the 
chiral Lagrangian approach by constraining the formfactor from comparison 
with results of a nonrelativistic approach which makes use of meson wave 
functions \cite{wsb00}. Finally, we will obtain a result for the 
off-shell $J/\psi$ breakup cross section which can be compared to the fit 
formula used in \cite{bbk}. This quantity is required for the calculation 
of the in-medium modification of the  $J/\psi$ breakup due to the 
Mott-effect for mesonic states at the deconfinement/chiral restoration 
transition which has been suggested \cite{bbk,bbk2} as an explanation of 
the anomalous $J/\psi$ suppression effect observed in heavy-ion 
collisions at the CERN-SPS \cite{na50}. 

\section{Effective Chiral Lagrangian} 
 
We start from QCD at low energy. The effective  chiral Lagrangian 
for pseudoscalar (Goldstone) mesons  can be written as 
 
\begin{eqnarray} 
{\cal L}_0 = \frac{F^2_\pi}{8} \mbox{tr} 
\left[\partial_\mu U (x) \partial_\mu U^+ (x)\right]\,, 
\end{eqnarray} 
with $F_\pi = 132 $~{MeV} being the weak pion decay constant, and 
$U (x) =\mbox{exp}\left[ 2i \varphi (x) /F_\pi\right]$. 
Notice that $U(x)$ transforms in a so-called non-linear representation of the 
$SU(N_f)_L \times SU(N_f)_R$ group. The usual 
multiplet of pseudoscalar mesons is $\varphi = \varphi^a \lambda^a /\sqrt{2}$ 
, $\lambda^a$ are Gell - Mann matrices. 
To introduce vector and axial-vector mesons we follow the procedure 
which is connected with the replacement 
\begin{eqnarray} 
{\cal L}_0 \longrightarrow {\cal L} = \frac{F^2_\pi}{8} \mbox{tr} 
\left[{\cal D}_\mu U {\cal D}_\mu U^+ \right]\,, 
\end{eqnarray} 
given by 
\begin{eqnarray} 
{\cal D}_\mu = \partial_\mu U -igA_\mu^L U + ig U A_\mu^R\,. 
\end{eqnarray} 
The left - and right- handed spin-1 fields, $A_\mu^L$ and $A_\mu^R$, 
are  combinations of vector and axial-vector meson fields 
\begin{eqnarray} 
A_\mu^L &=& \frac{1}{2} \left( V_\mu + A_\mu\right)\,, 
\nonumber \\ 
A_\mu^R &=& \frac{1}{2} \left( V_\mu  -  A_\mu\right)\,. 
\end{eqnarray} 
The coupling of these mesons to pseudoscalars is introduced as 
a gauge coupling with the gauge coupling constant $g$ which can be determined 
from the $\rho \longrightarrow \pi \pi$~ decay; $g_{\rho\pi\pi}=8.6$~. 
Therefore, the Lagrangian involving spin-1 and spin-0 mesons takes the form 
\begin{eqnarray} 
{\cal L}(\varphi,V,A)&=& 
{1\over 8}F_\pi^2 \mbox{tr}\left[ {\cal D}_\mu U({\cal D}_\mu U)^+ \right] 
+ {1\over 8} F_\pi^2 \mbox{tr} \left[ M(U+U^+ - 2) \right] 
\nonumber \\ && 
 -{1\over 2} \mbox{tr} \left[ (F_{\mu\nu}^L)^2 + (F_{\mu\nu}^R)^2 \right] 
+ m_0^2 \mbox{tr} \left[ (A_\mu^L)^2 + (A_\mu^R)^2 \right] 
\nonumber \\ && 
-i \xi \mbox{tr} \left[ ({\cal D}_\mu U)( {\cal D}_\nu U)^+ F_{\mu\nu}^L 
+ (D_\mu U)^+ (D_\nu U) F_{\mu\nu}^R   \right] 
\nonumber \\ && 
+\gamma \mbox{tr} \left[ F_{\mu\nu}^L U F_{\mu\nu}^R U^+ \right]\,. 
\label{Lagr} 
\end{eqnarray} 
The second term is proportional to the mass matrix $M$ and describes the 
``soft'' 
breaking of the chiral $SU(N_f)_L\times SU(N_f)_R$ symmetry. 
The corresponding field strength tensors are given by 
\begin{eqnarray} 
F_{\mu \nu}^{L,R} = 
\partial_\mu A_\nu^{L,R} -  \partial_\nu A_\mu^{L,R} 
-i g \left[A_\mu^{L,R}, A_\nu^{L,R} \right]~. 
\end{eqnarray} 
The third and fourth terms in (5) correspond to the free Lagrangians of the 
spin-1 particles. At this level of the chiral symmetry all spin- 1 mesons have 
the same ``bare'' mass,~$m_0$. The remaining terms in (5) are so - called 
non - minimal terms 
since they  contain higher orders in the derivatives 
as well as the mixed term ($\partial_\mu\varphi A_\mu $). After the 
diagonalization of (5) we obtain the  Lagrangians  
with pseudoscalar, vector and axial - vector mesons 
(see Appendix). 
The Lagrangian (\ref{Lagr}) contains many degrees of freedom. 
We can consider the special case when vector mesons are 
desribed as dynamical gauge bosons. It corresponds to 
 the ``hidden'' chiral symmetry. 
 We choose a gauge where 
left- and right-handed fields in the Lagrangian will be identical to the 
vector field \,$V_\mu$:\, 
$A_\mu^{L'}=A_\mu^{R'}=V_\mu$\, and \,$A'_{\mu}= 0.$ 
This can be achieved  by a gauge 
transformation which conserves the $SU(N_f)_L\times SU(N_f)_R$ symmetry 
\begin{eqnarray} 
A_\mu^L=A_\mu^R=V_\mu~, 
\nonumber \\ && 
U\longrightarrow U_L\ U\ U_R^+~, 
\nonumber \\ && 
A_\mu^L\longrightarrow U_L\ A_\mu^L\ U_L^+\ 
+\ {i \over g}\ U_L\ \partial_\mu U_R^+~, 
\nonumber \\ && 
A_\mu^L\longrightarrow U_R\ A_\mu^L\ U_R^+\ 
+\ {i \over g}\ U_R\ \partial_\mu U_R^+~, 
\end{eqnarray} 
with the specific choice $U_L=U^{1 \over 2}$ and $U_R=U^{-{1 \over 2}}$, 
so that pseudoscalar mesons are gauge parameters. 
Now we can rewrite the Lagrangian (5) as a sum of three Lagrangians 
 
\begin{eqnarray} 
{\cal L}_0&=& \frac{F^2_\pi}{8} \mbox{tr} 
\left({\cal D}_\mu U {\cal D}_\mu U^+ \right)~, 
\end{eqnarray} 
 
\begin{eqnarray} 
{\cal L}_1&=& -{1\over 2} \mbox{tr} \left( (F_{\mu\nu}^L)^2+ 
(F_{\mu\nu}^R)^2\right) 
+\gamma\, \mbox{tr} \left( F_{\mu\nu}^L U F_{\mu\nu}^R U^+ \right)~, 
\end{eqnarray} 
 
\begin{eqnarray} 
{\cal L}_2&=&m_0^2 \mbox{tr} \left( (A_\mu^L)^2+ (A_\mu^R)^2\right) 
+B\,\mbox{tr}\left( A_\mu^L U A_\mu^R U^+ \right) 
\nonumber \\ && 
+C\,\mbox{tr}\left( A_\mu^L A^{R\mu} + A_\mu^R A^{L\mu}\right)~. 
\label{Lag3} 
\end{eqnarray} 
 
Note that we have added two gauge invariant terms to the  Lagrangian (5). 
The second term (with the coefficient B) in (10) plays an important role 
in the description of the width of the $\rho\rightarrow \pi\pi$~decay, 
and the third term (with the coefficient C) maintains the gauge invariance 
of the J/$\psi+\pi\rightarrow D^*+\bar{D}$~decay. Applying the 
substitutions (7) to the Lagrangian (5), we obtain 
\begin{eqnarray*} 
{\cal L}_0\rightarrow {\cal L'}_0=0~, 
\end{eqnarray*} 
\begin{eqnarray*} 
{\cal L}_1\rightarrow {\cal L'}_1= 
(\gamma - 1)\,\mbox{tr}\left(F_{\mu\nu}^V\,F^{\mu\nu V}\right)~, 
\end{eqnarray*} 
 
\begin{eqnarray} 
{\cal L}_2\rightarrow {\cal L'}_2&=& 
\frac {m_V^2} {2} \mbox{tr}\left(V_\mu^2\right) 
- i{g_{V\varphi\varphi} \over 2}\mbox{tr}\left(V_\mu\left( \varphi 
\stackrel{\leftrightarrow}{\partial}_\mu 
\varphi \right)\right) 
\nonumber \\ 
&+& {8C \over F_\pi^2}\mbox{tr}\left(\left(V_\mu\varphi\right)^2- 
V_\mu^2\varphi^2\right)+{\cal L}(\varphi)~, 
\end{eqnarray} 
where the vector mass and the vector-pseudoscalar-pseudoscalar 
coupling are defined
by~$m_V^2=2(B+2m_0^2+2C),~g_{V\varphi\varphi}=2(B-2C+2m_0^2)/(gF_\pi^2)$~ .

\subsection{$J/\psi$ breakup cross sections} 
 
\begin{figure}[htb] 
\parbox{0.45\textwidth}{ 
\center{\epsfig{figure=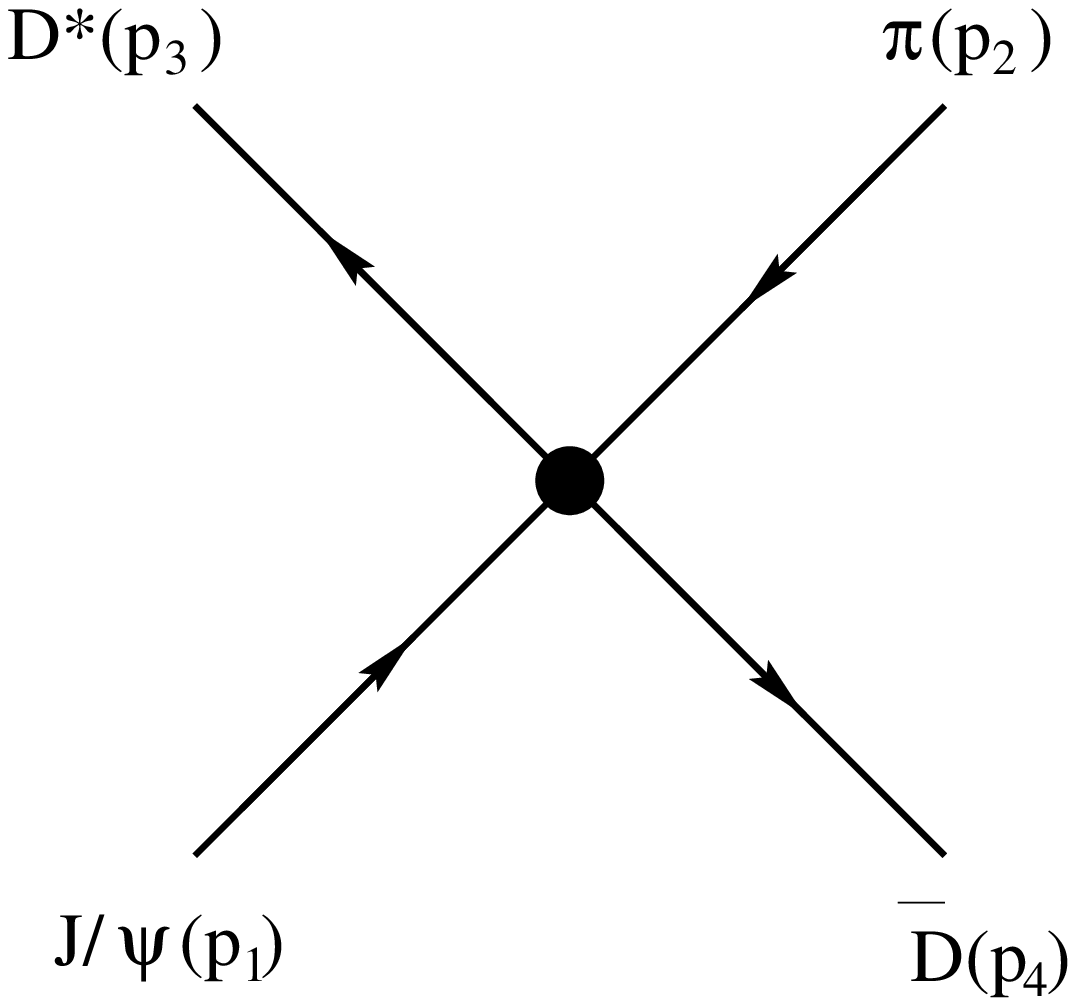,width=0.4\textwidth,angle=-90} 
\\{\bf I}} 
} 
\parbox{0.45\textwidth}{ 
\center{\epsfig{figure=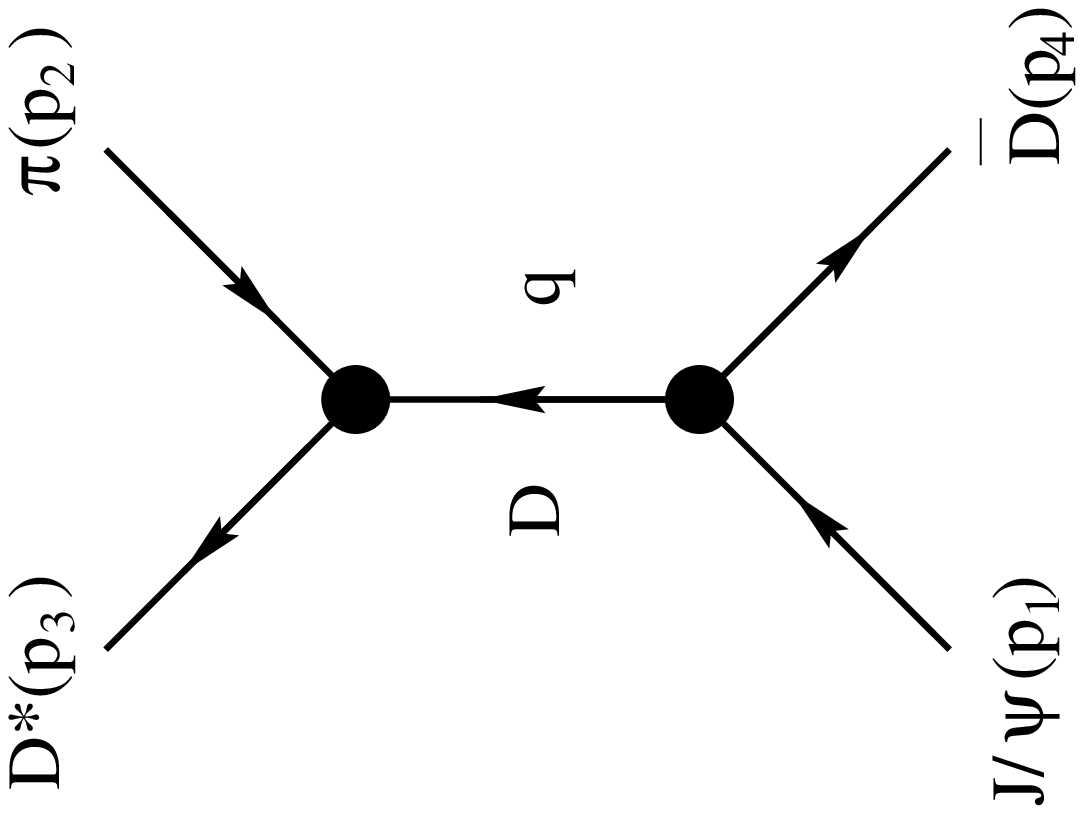,width=0.4\textwidth,angle=-90} 
\\{\bf III}} 
} \vspace{5mm} 
\center{\epsfig{figure=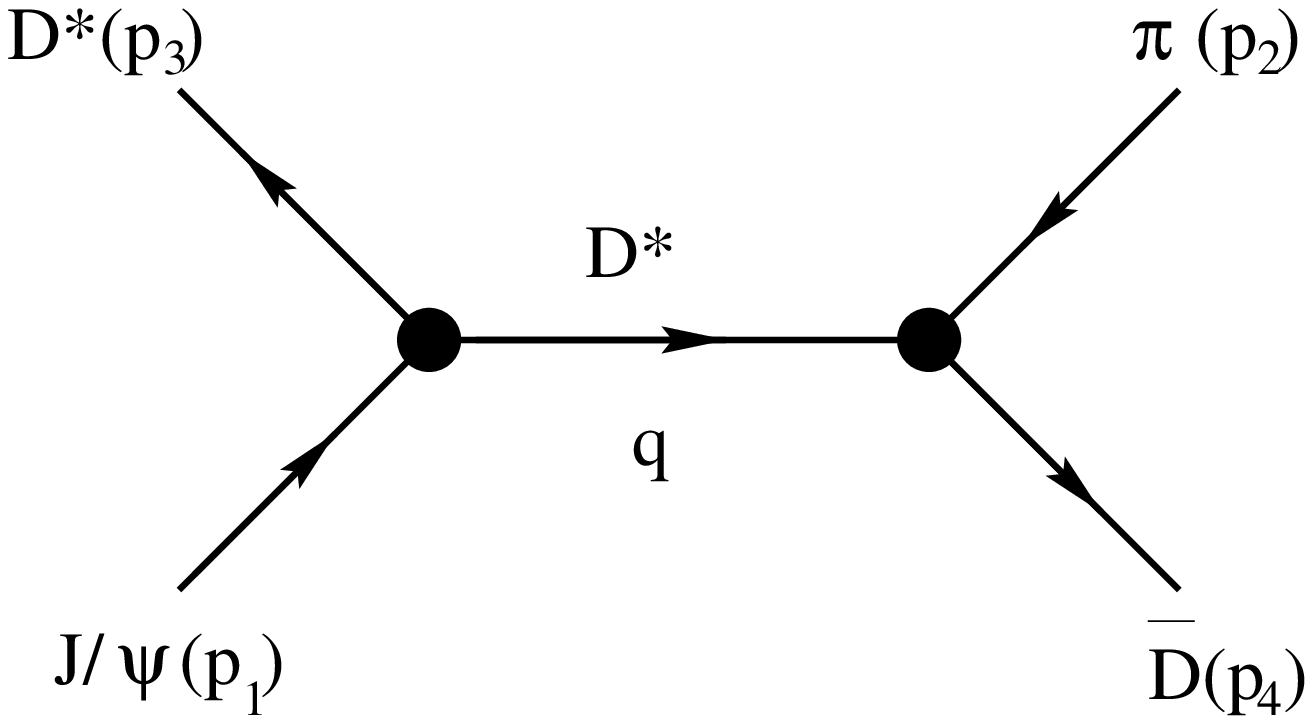,width=0.35\textwidth,angle=-90} 
\\{\bf II}}
\vspace{5mm} \caption{Diagrams for $J/\psi$ breakup by pion impact: 
$J/\psi+\pi\rightarrow D^*+\bar{D}$; I - contact term, II+III - D-meson 
exchange processes.}
\label{fig1}
\end{figure}

The above effective Lagrangian allows us to study the following processes for 
{$J/\psi$} breakup by $\pi$ and $\rho$ mesons 
\begin{eqnarray} 
&& J/\psi + \pi \rightarrow D^* + \bar{D},\ 
J/\psi + \pi \rightarrow D + \bar{D}^*,\ \\ 
&& J/\psi + \rho \rightarrow D + \bar{D},\ 
J/\psi + \rho \rightarrow D^* + \bar{D^*}~. 
\end{eqnarray} 
The generic diagrams for these processes are 
shown in Fig. \ref{fig1} for the example of the first one. 

The full amplitude for the first process 
{$J/\psi + \pi \longrightarrow D^* + \bar{D}$},~without isospin factors and 
before summing and averaging over external spins,~is given by 
\begin{eqnarray} 
{\cal M}_1\equiv{\cal M}_1^{\mu\nu}\varepsilon_{1\mu}\varepsilon_{3\nu}&=&
\left(\sum_{i=a,b,c} {\cal M}_{1i}^{\mu\nu}\right)\varepsilon_{1\mu}
\varepsilon_{3\nu}~, 
\end{eqnarray} 

with 
\begin{eqnarray*} 
{\cal M}_{1a}^{\mu\nu}= -g_{\pi D D^*}g_{J/\psi DD}(-2p_2+p_3)^\nu
\left({1\over u-m_D^2}\right)(p_2-p_3+p_4)^\mu ~, 
\end{eqnarray*} 
\begin{eqnarray*} 
{\cal M}_{1b}^{\mu\nu} &=& g_{\pi D D^*}g_{J/\psi D^* D^*} (-p_2 -p_4)^\alpha 
\left( {1 \over t-m_{D^*}^2}\right) 
\nonumber \\ 
&\times&\left[ g^{\alpha\beta} - {(p_2 -p_4)^\alpha (p_2 -p_4)^\beta 
\over m_{D^*}^2}\right] 
\nonumber \\ 
&\times&\left[(-p_1-p_3)^\beta g^{\mu\nu}+(-p_2+p_1+p_4)^\nu 
g^{\beta\mu}+(p_2+p_3-p_4)^\mu g^{\beta\nu}\right]\ , 
\end{eqnarray*} 
\begin{eqnarray} 
{\cal M}_{1c}^{\mu\nu} = -g_{J/\psi \pi D D^*}\  g^{\mu\nu} \ . 
\end{eqnarray} 
Similarly, the full amplitude for the second  process  
$J/\psi + \rho \rightarrow D + \bar{D}$ is given by 
\begin{eqnarray} 
{\cal M}_2\equiv{\cal M}_2^{\mu\nu}\varepsilon_{1\mu}\varepsilon_{2\nu}&=&
\left(\sum_{i=a,b,c} {\cal M}_{2i}^{\mu\nu}\right)\varepsilon_{1\mu}
\varepsilon_{2\nu} 
\end{eqnarray} 
with 
\begin{eqnarray} 
\nonumber \\ && 
{\cal M}_{2a}^{\mu\nu} = -g_{\rho D D}g_{J/\psi DD}(p_2-2p_3)^\mu
\left({1\over u-m_D^2}\right)(p_2-p_3+p_4)^\nu \ , 
\nonumber \\ && 
{\cal M}_{2b}^{\mu\nu} = -g_{\rho D D}g_{J/\psi DD}(-p_2+2p_4)^\mu
\left({1\over t-m_D^2}\right)(-p_2-p_3+p_4)^\nu \ , 
\nonumber \\ && 
{\cal M}_{2c}^{\mu\nu} = g_{J/\psi \rho D D}\  g^{\mu\nu} \ . 
\end{eqnarray} 
 
For the third process $J/\psi + \rho \rightarrow D^* + \bar{D^*}$,~the full 
amplitude is given by

\begin{eqnarray} 
{\cal M}_3\equiv{\cal M}_3^{\mu\nu\lambda\omega}\varepsilon_{1\mu}
\varepsilon_{2\nu}\varepsilon_{3\lambda}\varepsilon_{4\omega}&=&
\left(\sum_{i=a,b,c} {\cal M}_{3i}^{\mu\nu\lambda\omega}\right)
\varepsilon_{1\mu}\varepsilon_{2\nu}\varepsilon_{3\lambda}
\varepsilon_{4\omega}\ , 
\end{eqnarray} 
\begin{eqnarray*} 
{\cal M}_{3a}^{\mu\nu\lambda\omega}&=& 
g_{\rho D^* D^*}g_{J/\psi D^* D^*}\left[(-p_2-p_3)^\alpha g^{\mu\lambda} 
+2p_2^\lambda g^{\alpha\mu}+2p_3^\mu g^{\alpha\lambda} \right] 
\nonumber \\ 
&\times& \left({1 \over u-m_{D^*}^2}\right) 
\left[g^{\alpha\beta}-{(p_2 -p_3)^\alpha (p_2-p_3)^\beta\over m_{D^*}^2}\right]
\nonumber \\ 
&\times& \left[-2p_1^\omega g^{\beta\nu}+(p_1+p_4)^\beta 
g^{\nu\omega}-2p_4^\nu g^{\beta\omega}\right]~, 
\end{eqnarray*} 
\begin{eqnarray*} 
{\cal M}_{3b}^{\mu\nu\lambda\omega}&=& 
g_{\rho D^* D^*}g_{J/\psi D^* D^*} 
\left[-2p_2^\omega g^{\alpha\mu}+(p_2+p_4)^\alpha g^{\mu\omega}-2p_4^\mu 
g^{\alpha\omega} \right] 
\nonumber \\ 
&\times&\left({1 \over t-m_{D^*}^2}\right) 
\left[ g^{\alpha\beta} - {(p_2 -p_4)^\alpha (p_2 -p_4)^\beta 
\over m_{D^*}^2}\right]\nonumber \\ 
&\times&\left[(-p_1-p_3)^\beta g^{\nu\lambda}+2p_1^\lambda 
g^{\beta\nu}+2p_3^\nu g^{\beta\lambda}\right]\ , 
\end{eqnarray*} 
\begin{eqnarray} 
{\cal M}_{3c}^{\mu\nu\lambda\omega}=g_{J/\psi\rho D^* D^*}(g^{\mu\lambda} 
g^{\nu\omega} + g^{\mu\omega} g^{\nu\lambda}-2 g^{\mu\nu} g^{\lambda\omega})\ .
\end{eqnarray}
In the above,~$p_j$ denotes the momentum of particle~$j$. 
We choose the convention that particle $1$ and $2$ represent 
initial-state mesons while particles $3$ and $4$ represent 
final-state mesons on the left and right sides of 
the diagrams shown in Fig. 1,~respectively. 
The indices $\mu,~\nu,~\lambda$ and $\omega$~ denote 
the polarization components of external particles while 
the indices $\alpha$ and $\beta$ denote those of the exchanged mesons. 
\par After averaging 
  (summing) over initial (final) spins and including isospin factors,~ 
the cross sections for the the three processes are given by 
 
\begin{eqnarray} 
{d\sigma_1 \over dt}= 
{1 \over {96\pi s p_{i,c.m.}^2} }{\cal M}_1^{\mu\nu} 
{\cal M}_1^{* \mu ' \nu '} 
\left( g^{\mu\mu '} - { p_1^\mu p_1^{\mu '} \over m_1^2 }\right) 
\left( g^{\nu\nu '} - { p_3^\nu p_3^{\nu '} \over m_3^2 }\right), 
\end{eqnarray} 
 
\begin{eqnarray} 
{d\sigma_2 \over dt} 
={1 \over 288\pi s p_{i,c.m.}^2} 
{\cal M}_2^{\mu\nu}{\cal M^*}_2^{\mu ' \nu '} 
\left(g^{\mu\mu '}-{p_1^\mu p_1^{\mu '} \over m_1^2}\right) 
\left(g^{\nu\nu '} - {p_2^\nu p_2^{\nu '} \over m_2^2 }\right), 
\end{eqnarray} 
 
\begin{eqnarray} 
{d\sigma_3 \over dt}&=& 
{1 \over 288\pi s p_{i,c.m.}^2} 
{\cal M}_3^{\mu\nu\lambda\omega} 
{\cal M^*}_3^{\mu ' \nu '\lambda '\omega '} 
\left(g^{\mu\mu '}-{p_1^\mu p_1^{\mu '} \over m_1^2}\right) 
 \left(g^{\nu\nu '} - {p_2^\nu p_2^{\nu '} \over m_2^2 }\right) 
\nonumber \\ 
&\times& 
\left(g^{\lambda\lambda '} - {p_3^\lambda p_3^{\lambda '} \over m_3^2 } 
\right) 
\left(g^{\omega\omega '} - {p_4^\omega p_4^{\omega '} \over m_4^2 }\right), 
\end{eqnarray} 
with $s=(p_1+p_2)^2$,~and 
\begin{eqnarray} 
p_{i,c.m.}^2={[s-(m_1+m_2)^2][s-(m_1-m_2)^2]\over 4s}~, 
\end{eqnarray} 
is the squared momentum of initial-state mesons in the center-of-momentum 
(c.m.) frame. The definition of $p_{f,c.m.}$ for the final-state mesons 
is analogous with the replacement $(m_1,m_2)\to(m_3,m_4)$. 
 
\section{Hadronic Formfactors}

\begin{figure}[htb]
\epsfig{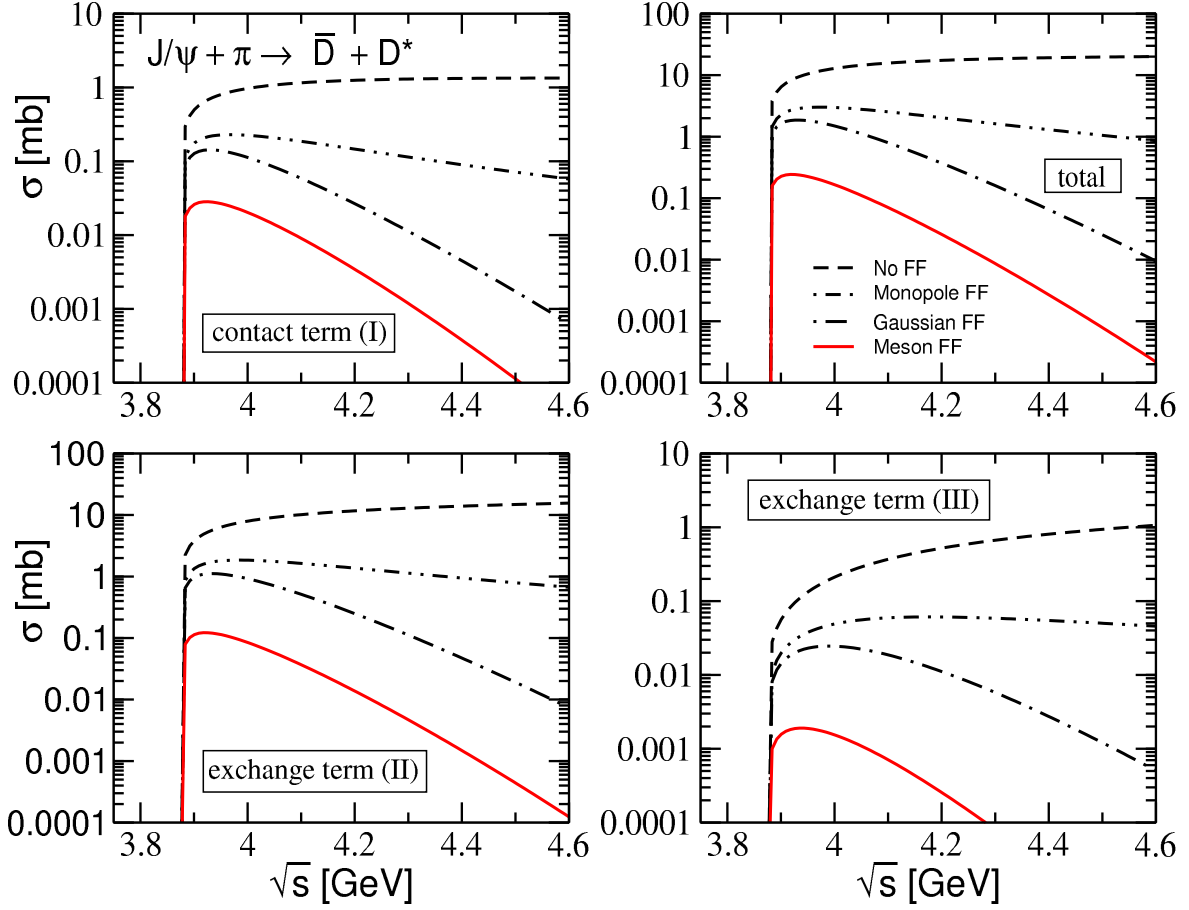}
\vspace{5mm} \caption{Upper right panel: total cross section for $J/\psi$ 
break-up by pion impact without formfactor (black dashed line), with monopole 
type formfactor (black dotted-dashed-dotted line), with Gaussian formfactor 
(black dotted-dash line) and with ``meson'' formfactor (red solid line) 
as a function of the c.m. energy of initial-state mesons. 
The partial contributions from the diagrams I, II, and III of Fig. \ref{fig1} 
are shown in the other panels.}
\label{fig2}
\end{figure}

The chiral Lagrangian aproach for $J/\psi$ breakup by light meson impact 
makes the assumption that mesons and meson-meson interaction vertices are 
pointlike (four-momentum independent) objects. This neglect of the finite 
extension of mesons as quark-antiquark bound states has dramatic 
consequences: it leads to a monotonously rising behaviour of the cross 
sections for the corresponding processes, see the dashed lines in Fig. 
\ref{fig2}. 
This result, however, cannot be correct away from the rection threshold 
where the tails of the mesonic wave functions determine the high-energy 
behaviour of the quark exchange (in the nonrelativistic formulation of 
\cite{mbq95,wsb00}) or quark loop (in the relativistic formulation 
\cite{b+00}) diagrams describing the microscopic processes underlying the 
$J/\psi$ breakup by meson impact. As long as the mesonic wave functions 
describe quark-antiquark bound states which have a finite extension in 
coordinate- and momentum space, the $J/\psi$ breakup cross section is 
expected to be decreasing above the reaction threshold and asymptotically 
small at high c.m. energies. This result of the quark model approaches to 
meson-meson interactions \cite{mbq95,wsb00,b+00} can be mimicked within 
chiral meson Lagrangian approaches by the use of formfactors at the 
interaction vertices \cite{lk00,hg00}.
 
\subsection{Global formfactor ansatz}

We will follow here the definitions 
of Ref. \cite{lk00}, where the formfactor of all the four-point vertices of 
Fig.1, i.e. that of the box diagram (I) as well as that of the meson exchange 
diagrams (II, III) is taken as a product of the triangle diagram 
formfactors 
\begin{equation} 
\label{f3f3} F_4^i({\bf q}^2)=\left[F_3({\bf 
q}^2)\right]^{2}~~,~i=I,II,III~, 
\end{equation} 
with the squared three-momentum ${\bf q}^2$ given by the average value of 
the squared three-momentum transfers in the $t$ and $u$ channels 
\begin{equation} 
{\bf q}^2=\frac{1}{2}\left[({\bf p_1}-{\bf p_3})^2+ ({\bf p_1}-{\bf 
p_4})^2\right]_{\rm c.m.}= p^2_{i,{\rm c.m.}}+p^2_{f,{\rm c.m.}}~. 
\end{equation} 
For the triangle diagrams, we use formfactors with a momentum dependence 
in the monopole form ($M$) 
\begin{equation} 
F_3^M({\bf q}^2)=\frac{\Lambda^2}{\Lambda^2 +{\bf q}^2 }~, 
\end{equation} 
and in the Gaussian ($G$) form 
\begin{equation} 
F_3^G({\bf q}^2)=\exp(-{{\bf q}^2/\Lambda^2})~. 
\end{equation} 

\subsection{Meson formfactor ansatz}

In order to take into account the quark substructure of meson-meson vertices 
we will suggest here a simple ansatz which respects the different size of the
interacting mesons and the different quark diagram representation of contact
and meson exchange interactions in terms of quark box and quark triangle 
diagrams. 
The triangle diagram is of third order in the wave functions so that the meson 
exchange diagrams are suppressed at large momentum transfer by six wave 
functions, the box diagram appears already at fourth order thus being 
less suppressed than suggested by the ansatz (\ref{f3f3}) of Ref. 
\cite{lk00}.

For the contact term $(I)$ we use the replacement 
$g_{J/\psi \pi D^* D} \longrightarrow 
g_{J/\psi \pi D^* D} \times F_I(s)\,$
where the formfactor has the following form 

\begin{eqnarray}
F_I(s) = \exp\Bigg\{ 
-\frac{1}{4 s}\Bigg[
\Big( s &-& (m_1 + m_2)^2 \Big) \Big( s - (m_1 - m_2)^2 \Big) 
\bigg( \frac{1}{\Lambda_1^2} + \frac{1}{\Lambda_2^2} \bigg) \nonumber \\
&+& \Big( s - (m_3 + m_4)^2 \Big) \Big( s - (m_3 - m_4)^2 \Big) 
\bigg( \frac{1}{\Lambda_3^2} + \frac{1}{\Lambda_4^2} \bigg)
\Bigg]
\Bigg\}\;
\end{eqnarray}

The second exchange term $(II)$ 
can be written using the substitution 
$g_{J/\psi D^* D^*} \times g_{D^* D \pi} 
\longrightarrow g_{J/\psi D^* D^*} \times g_{D^* D \pi} \times F_{II}(s,t)\,$
with
\begin{eqnarray}
F_{II}(s,t) = \exp\Bigg\{ 
-\frac{1}{4 s}\Bigg[
\Big( s &-& (m_1 + m_2)^2 \Big) \Big( s - (m_1 - m_2)^2 \Big) 
\bigg( \frac{1}{\Lambda_1^2} + \frac{1}{\Lambda_2^2} \bigg) \nonumber \\
&+& \Big( s - (m_3 + m_4)^2 \Big) \Big( s - (m_3 - m_4)^2 \Big) 
\bigg( \frac{1}{\Lambda_3^2} + \frac{1}{\Lambda_4^2} \bigg) \nonumber \\
&&\qquad + \frac{2}{\Lambda_4^2} \bigg[ 
\bigg( \Big( m_1^2 - m_2^2 \Big) - \Big( m_3^2 - m_4^2 \Big) \bigg)^2 
- 4 s t \bigg]
\Bigg]
\Bigg\}\;
\end{eqnarray}
Analogously,  
the exchange term $(III)$ is obtained by 
 $g_{J/\psi D D} \times g_{D^* D \pi} 
\longrightarrow g_{J/\psi D D} \times g_{D^* D \pi} \times F_{III}(s,u)\,$
with the formfactor
\begin{eqnarray}
F_{III}(s,u) = \exp\Bigg\{ 
-\frac{1}{4 s}\Bigg[
\Big( s &-& (m_1 + m_2)^2 \Big) \Big( s - (m_1 - m_2)^2 \Big) 
\bigg( \frac{1}{\Lambda_1^2} + \frac{1}{\Lambda_2^2} \bigg) \nonumber \\
&+& \Big( s - (m_3 + m_4)^2 \Big) \Big( s - (m_3 - m_4)^2 \Big) 
\bigg( \frac{1}{\Lambda_3^2} + \frac{1}{\Lambda_4^2} \bigg) \nonumber \\
&&\qquad + \frac{2}{\Lambda_3^2} \bigg[ 
\bigg( \Big( m_1^2 - m_2^2 \Big) + \Big( m_3^2 - m_4^2 \Big) \bigg)^2 
- 4 s u \bigg]
\Bigg]
\Bigg\}\;
\end{eqnarray}
Formfactors depend on the 
parameters which we fix from the physical observables 
(decay widths $\rho \longrightarrow \pi \pi $, 
$D^* \longrightarrow D \pi$) and the vector dominance model 
\cite{lk00,hg00}.
We use for the coupling constants the values 
$g_{D^* D \pi} = g_{D D \varrho} = g_{D^* D^* \varrho} = 4.4$, 
$g_{J/\psi D D} = g_{J/\psi D^* D^*} = 7.7$, 
$g_{J/\psi \pi D^* D} = g_{J/\psi \varrho D^* D} = g_{J/\psi \varrho D D} 
= 33.9$.
For the range parameters $\Lambda_i$ of the quark-antiquark-meson vertices
we suggest to use the meson masses $m_i$, see Table \ref{table}.
The results are depicted in Fig. \ref{fig2}. 
In the last Section, we discuss the results and their possible 
implications for phenomenological applications. 

\section{Results and Discussion} 
 
The $J/\psi$ breakup cross section by $\pi$ and $\rho$ meson impact has 
been formulated within a chiral $U(4)$ Lagrangian approach. Numerical 
results have been obtained for the pion impact processes with the result 
that the D-meson exchange in the t-channel is the dominant subprocess 
contributing to the $J/\psi$ breakup. The use of formfactors at the 
meson-meson vertices is mandatory since otherwise the high-energy 
asymptotics of the processes with hadronic final states will be 
overestimated, see Fig. \ref{fig2}. 
From a comparison of the monopole and Gaussian functions in the global 
formfactor ansatz, we observe a big difference in the corresponding cross 
sections above the threshold. Generally we would prefer the use of gaussian 
functions which are motivated by strong (confining) quark-antiquark 
interactions within the mesons.
The new meson formfactor scheme reduces the peak value of the $J/\psi$ 
breakup cross section relative to the global scheme by 50 $\% $. 
The net result for the considered pion impact subprocess is in good 
correspondence to the one from the nonrelativistic quark exchange model.

Finally, we want to present an exploratory study of the influence of 
a variation of the 
final state D-meson masses on the effective $J/\psi$ breakup cross 
section. Our motivation for considering mesonic states to be off their 
mass-shell is their compositeness which can become apparent in a 
high-temperature (and density) environment at the deconfinement/chiral 
restoration transition, when these states change their character 
qualitatively being resonant quark- antiquark scattering states in the 
quark plasma rather than on-shell mesonic bound states.

The consequence 
of this Mott-transition from bound to resonant states for the $J/\psi$ 
breakup has been explored by Burau et al. \cite{bbk,bbk2}, 
using a fit formula for the D-mass dependence of the breakup 
cross section which shows a strong enhancement when the process becomes 
subthreshold. This behaviour is qualitatively 
approved within the present chiral U(4) Lagrangian + formfactor model 
although the subthreshold enhancement is more moderate, see Fig. 
\ref{fig4}. 
A more consistent description should include a quark model derivation of 
the formfactors for the meson-meson vertices and their possible medium 
dependence. Such an investigation is in progress. 
 
\subsection*{Acknowledgement} 
V.I. and Yu.K. acknowledge support from the Deutsche 
Forschungsgemeinschaft under grant no. 436 RUS 17/102/00 and from the 
Ministery for Education, Science and Culture of Mecklenburg-Western Pommerania.

\appendix

\begin{appendix}

\section{Chiral Langrangians}
In this Appendix we write the explicit form of the chiral Langrangians.
We use these Langrangians for the calculations  of matrix elements
of the $J/\psi$ breakup cross section by $\pi$ and $\rho$ mesons

\begin{eqnarray} 
{\cal L}^{(2)}(\varphi,V,A)&=&{1 \over 2}\mbox{tr}(\partial_\mu \varphi)^2 
-{1 \over 2}\mbox{tr}\left(M\varphi^2\right) 
-{1\over 4}\mbox{tr}(F_{\mu\nu}^V)^2 
+{1\over 2}m_V^2\mbox{tr}\left(V_\mu\right)^2\nonumber\\ 
&&-{1\over 4}\mbox{tr}(F_{\mu\nu}^A)^2 
+{1\over 2}m_A^2\mbox{tr}\left(A_\mu\right)^2~, 
\end{eqnarray} 
 
\begin{eqnarray} 
{\cal L}(\varphi^4)&=& 
-{2\over 3F_\pi^2}\mbox{tr}(\partial_\mu\varphi\partial_\mu(\varphi^3)) 
+{1\over 2F_\pi^2}Z^2\mbox{tr}(\partial_\mu(\varphi^2)\partial_\mu(\varphi^2)) 
\nonumber\\ 
&& +{g^2\over 4m_V^2}\mbox{tr}(\partial_\mu\varphi 
\{\varphi^2,\partial_\mu\varphi\}) 
+{1\over 6F_\pi^2}\mbox{tr}\large(M\varphi^4\large) 
\\ && 
+\left( {1\over 8}g^2(1-\gamma)^2\alpha^4 
-\xi {2g \over F_\pi^2}Z^4\alpha^2\sqrt{1-\gamma}\right)\mbox{tr}
(\partial_\mu\varphi\partial_\nu\varphi[\partial_\mu\varphi,
\partial_\nu\varphi])~,\nonumber 
\end{eqnarray} 
 
\begin{eqnarray} 
{\cal L}(VV\varphi\varphi)&=& 
-{g^2\over 4Z^4}\mbox{tr}\left((V_\mu\varphi)^2-V_\mu^2\varphi^2\right) 
\nonumber \\ && 
-{1\over F_\pi^2}{\gamma\over {1-\gamma}} 
\mbox{tr}\left(\varphi^2(F_{\mu\nu}^V)^2-(F_{\mu\nu}^V\varphi)^2\right) 
\nonumber \\ && 
+{1\over 16}g^2\alpha^2(1+\gamma)\mbox{tr} 
\left(\large[\partial_\mu\varphi,V_\nu\large] 
+\large[V_\mu,\partial_\nu\varphi\large]\right)^2 
\nonumber \\ && 
+{1\over 8}g^2\alpha^2(1-\gamma)\mbox{tr} 
\left(\left[V_\mu,V_\nu\right][\partial_\mu\varphi,\partial_\nu\varphi]\right) 
\nonumber \\ && 
+{g\alpha \over 2F_\pi}{\gamma \over \sqrt{1-\gamma}}\mbox{tr} 
\left(\varphi[F_{\mu\nu}^V,\left([\partial_\mu\varphi,V_\nu] 
+[V_\mu,\partial_\nu\varphi]\right)]\right) 
\nonumber \\ && 
-{2g\xi\over F_\pi^2}{Z^4\over \sqrt{1-\gamma}}\mbox{tr} 
\left(\partial_\mu\varphi\partial_\nu\varphi[V_\mu,V_\nu]\right) 
\nonumber \\ && 
+{2g\xi\over F_\pi^2}{Z^2 \over \sqrt{1-\gamma}}\mbox{tr} 
\left( (\partial_\mu\varphi[\varphi,V_\nu] 
+[\varphi,V_\mu]\partial_\nu\varphi)F_{\mu\nu}^V\right)~, 
\end{eqnarray} 

\begin{eqnarray} 
{\cal L}(A,V,\varphi)&=&-i{g^2F_\pi\over 4Z^2}\sqrt{{1-\gamma}
\over{1+\gamma}}\mbox{tr}\left(V_\mu[A_\mu,\varphi]\right) 
\nonumber \\ && 
+i{1\over F_\pi}{\gamma\over \sqrt{1-\gamma^2}}\mbox{tr}
\left(\varphi[F_{\mu\nu}^V,F_{\mu\nu}^A]\right) 
\nonumber \\ && 
+i{g^2 F_\pi\over 4m_V^2Z^2}(1-\delta)\sqrt{{1-\gamma}
\over{1+\gamma}}\mbox{tr}\left(F_{\mu\nu}^V[A_\mu,\partial_\nu\varphi]\right) 
\nonumber \\ && 
+i{g^2 F_\pi\over 4m_V^2}\sqrt{{1+\gamma}\over {1-\gamma}}\mbox{tr}
\left(F_{\mu\nu}^A[V_\mu,\partial_\nu\varphi]\right)~, 
\end{eqnarray} 
\begin{eqnarray} 
{\cal L}(V\varphi\varphi)&=& 
-i{g\over 2}\mbox{tr}\left(V_\mu 
\left( \varphi 
\stackrel{\leftrightarrow}{\partial}_\mu 
\varphi \right) 
\right) 
+i{g\delta\over 2m_V^2}\mbox{tr}\left(F_{\mu\nu}^V\partial_\mu\varphi
\partial_\nu\varphi\right)~, 
\end{eqnarray} 
\begin{eqnarray} 
{\cal L}(V^4)&=&{1 \over 16}{g^2\over 1-\gamma}\mbox{tr}\left([V_\mu,V_\nu]^2
\right)~, 
\end{eqnarray} 
\begin{eqnarray} 
{\cal L}(V^3)&=&i{g \over 4}\mbox{tr}\left( F_{\mu\nu}[V_\mu,V_\nu] \right)~, 
\end{eqnarray} 
where 
$$ 
\delta=1-Z^2-{2Z^4\over 1-Z^2}{\xi g \over \sqrt{1-\gamma}}~, 
$$ 
and $F_{\mu\nu}^V=\partial_\mu V_\nu - \partial_\nu V_\mu,\, 
F_{\mu\nu}^A=\partial_\mu A_\nu - \partial_\nu A_\mu,\, 
(\varphi\stackrel{\leftrightarrow}{\partial}_\mu\varphi)=\varphi\,
\partial_\mu\varphi-\partial_\mu\varphi\ \varphi$, which have the standard 
definition for commutators and anticommutators 
\cite{iz}.

\end{appendix}

\begin{table}\label{table}
 \begin{tabular}[h]{|c|c|c|c|c|c|}
  \hline
  {state $i$} & $J/\psi~ $ & $D^*~ $ & $D~ $ & $\varrho~ $ & $\pi~ $ \\
  \hline \hline
  {$m_i [{\rm GeV}]$} & 3.1 & 2.01 & 1.87 & 0.77 & 0.14 \\
  \hline
  {$\Lambda_i [{\rm GeV}]$} & 3.1 & 2.0 & 1.9 & 0.8 & 0.6 \\
  \hline
 \end{tabular}
\vspace{1cm}
\caption{Meson masses and range parameters coresponding to the 
quark-antiquark-meson vertices.}
\end{table}

\begin{figure}[tb] 
\epsfig{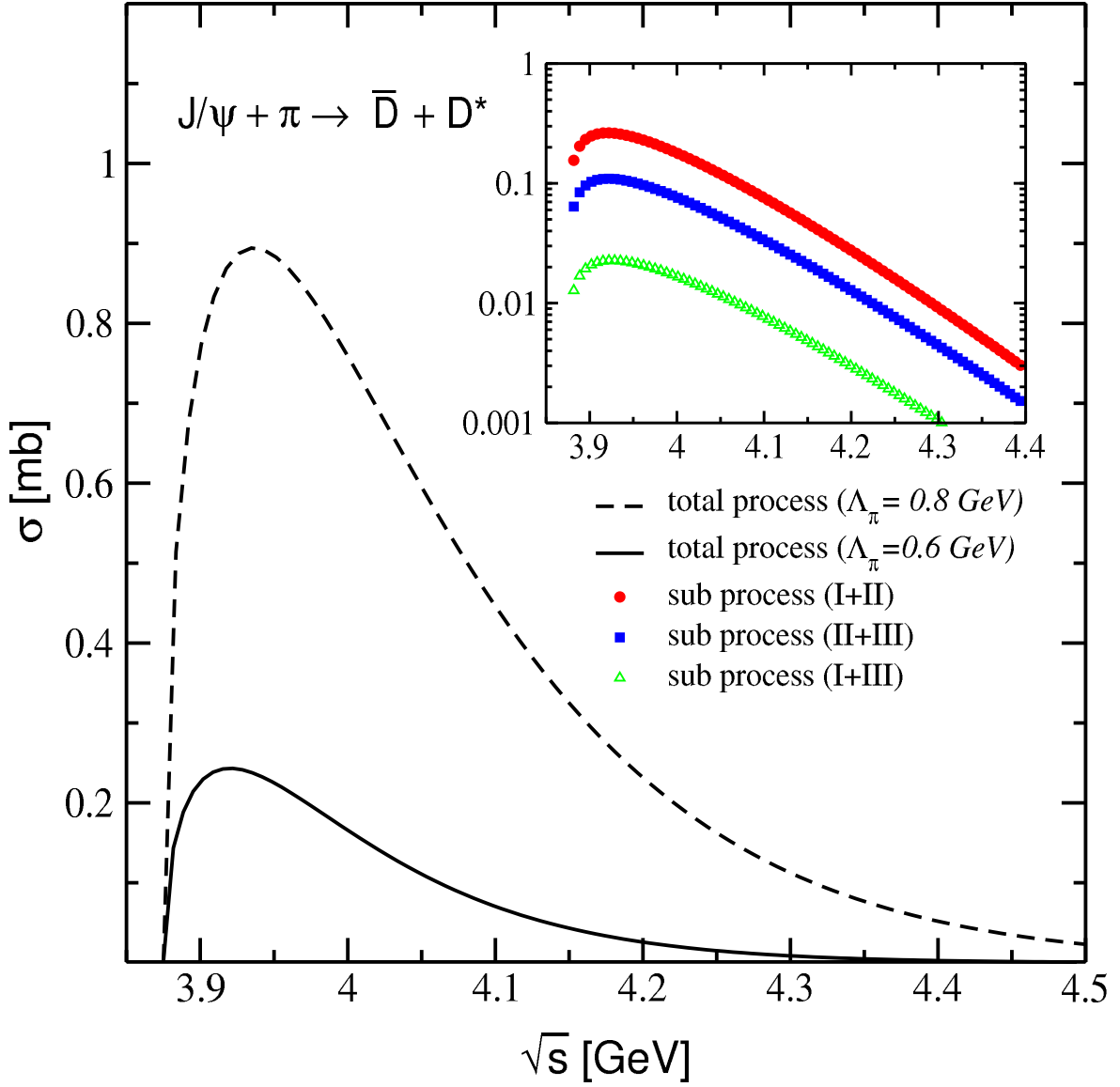} \vspace{5mm} 
\caption{Total $J/\psi + \pi \longrightarrow \bar{D} + D^*$ cross section 
with our meson formfactor (black solid line) in comparison with the 
cross sections of all sub-processes (contact term, exchange terms and 
corresponding interference terms)}
\label{fig3}
\end{figure}

\begin{figure}[htb]
\epsfig{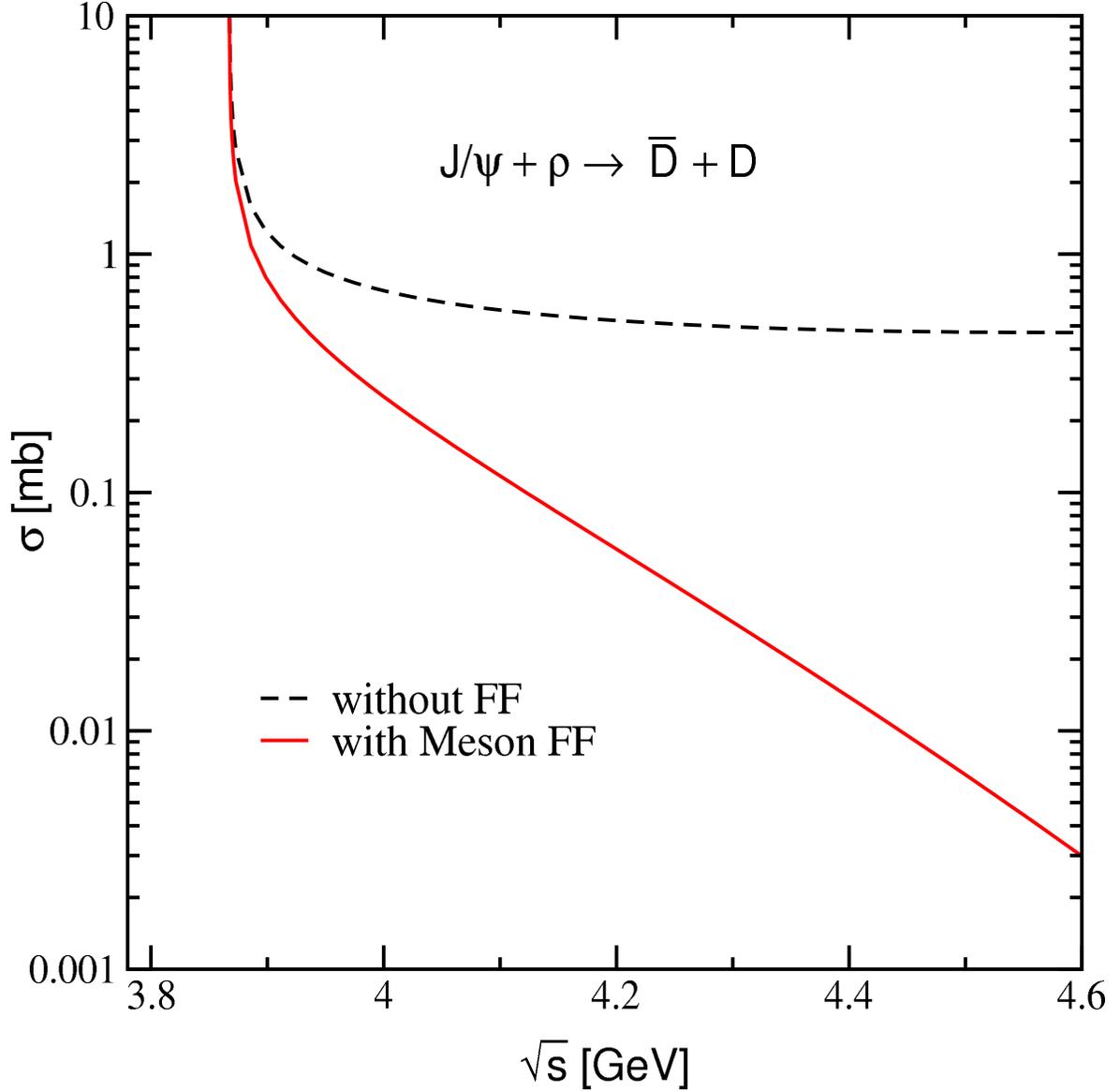}
\vspace{5mm} \caption{Total cross section for $J/\psi$ break-up by 
$\varrho$-meson impact due to the exothermic process 
$J/\psi + \rho \rightarrow \bar{D} + D$ without formfactor 
(black dashed line) and with meson formfactor (red solid line) as a function 
of the c.m. energy of initial-state mesons.}
\label{fig4}
\end{figure}

\begin{figure}[htb]
\epsfig{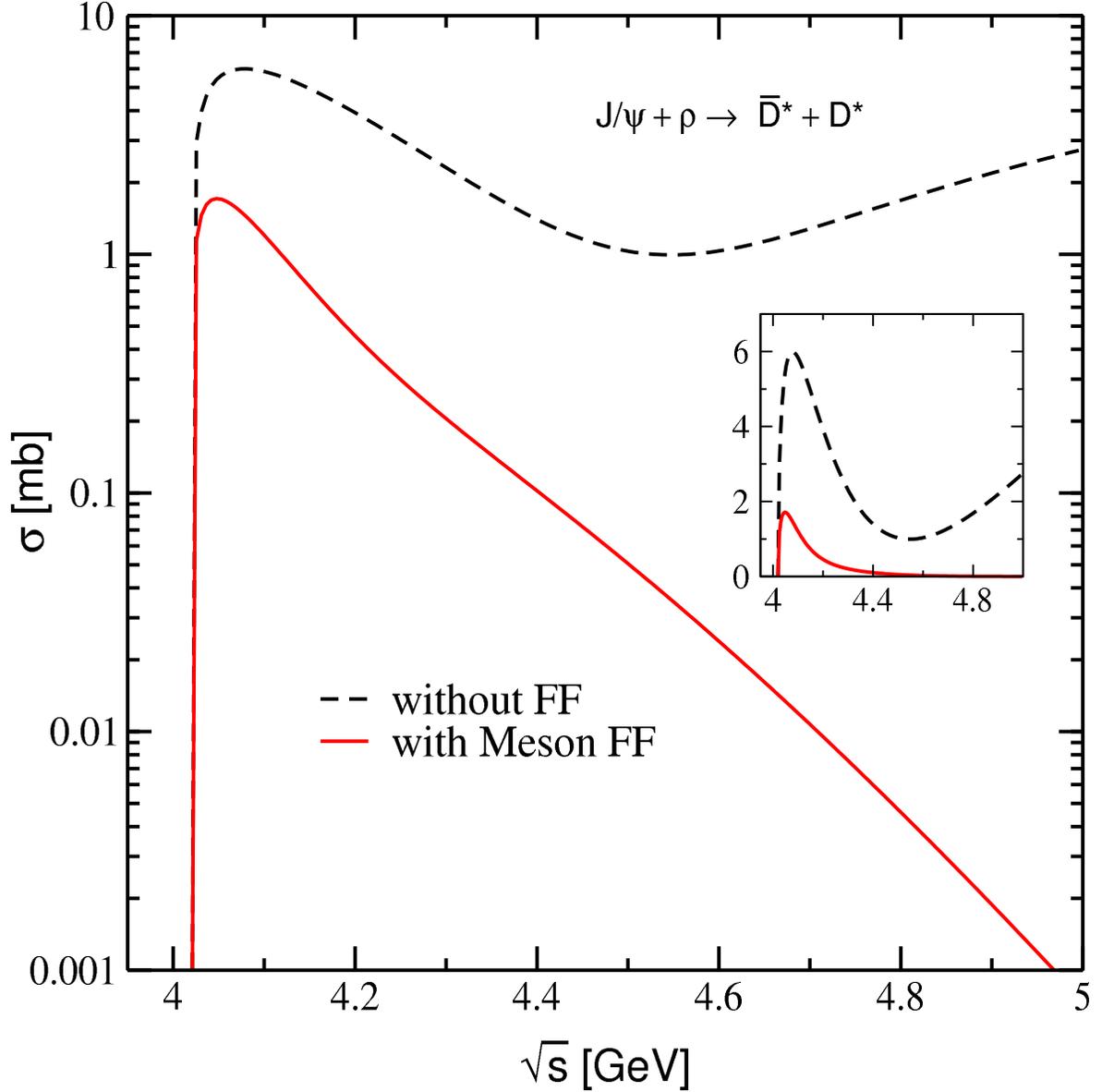}
\vspace{5mm} \caption{The same like in Fig. \ref{fig4}, but due to the 
endothermic process $J/\psi + \rho \rightarrow \bar{D^*} + D^*$ 
without formfactor (black dashed line) and with meson formfactor 
(red solid line) as a function of the c.m. energy of initial-state mesons. 
A representation with linear scaling is shown in the small panel.}
\label{fig5}
\end{figure}

\begin{figure}[htb]
\epsfig{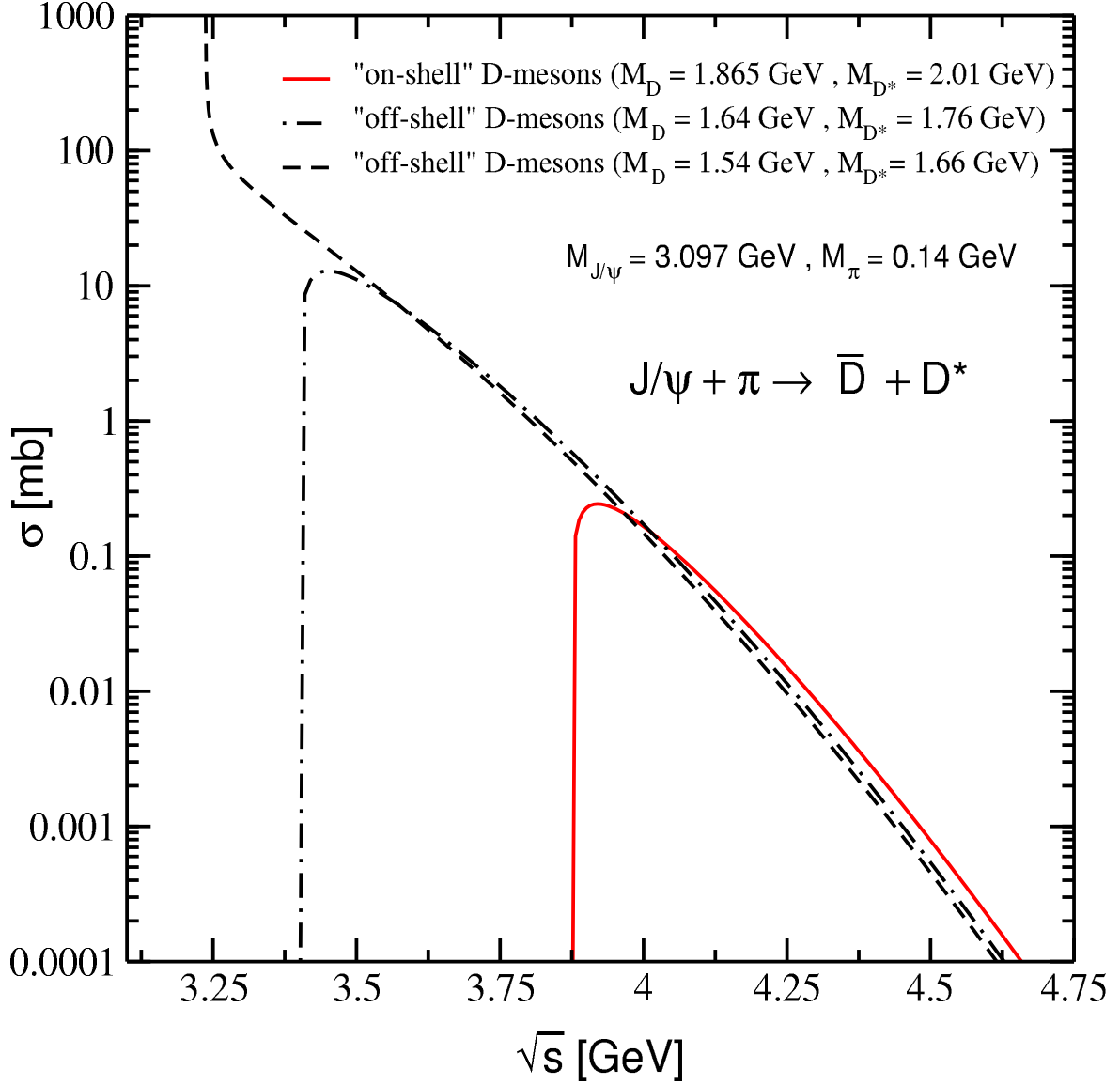}
\vspace{5mm} \caption{Total $J/\psi$ break-up cross section in the chiral 
Lagrangian model with mesonic formfactor when the final state masses 
$M_{D_1}=M_{D_2}=M_{D}$ are varied (``off-shell'') in the way that the 
reaction threshold for the process 
$J/\psi + \pi \longrightarrow \bar{D} + D^*$ 
is decreasing (endothermic; black dotted-dashed line) 
and finally vanishing (exothermic; black dashed line). 
The red solid line shows the ``on-shell'' result (the same like in Fig. 
\ref{fig3}).}
\label{fig6}
\end{figure}
 
\end{document}